\documentclass[aps,prl,twocolumn,superscriptaddress]{revtex4}
\usepackage{graphicx }
\newcommand{\ten}[1]{\stackrel{\leftrightarrow }{\bf #1}\! \! }
\begin{document}

\title{Fr\"ohlich mass in GaAs-based structures}

\author{C. Faugeras}

\affiliation{Grenoble high Magnetic Field laboratory, MPI-FKF and CNRS, B.P. 166, 38042 Grenoble Cedex 9, France}
\author{G. Martinez}
\affiliation{Grenoble high Magnetic Field laboratory, MPI-FKF and CNRS, B.P. 166, 38042 Grenoble Cedex 9, France}
\author{A. Riedel}
\affiliation{Paul Drude Institute, Hausvogteiplatz 5-7, D-10117 Berlin, Germany}
\author{ R. Hey }
\affiliation{Paul Drude Institute, Hausvogteiplatz 5-7, D-10117 Berlin, Germany}
\author{ K. J. Friedland}
\affiliation{Paul Drude Institute, Hausvogteiplatz 5-7, D-10117 Berlin, Germany}
\author{Yu. Bychkov }
\affiliation{Grenoble high Magnetic Field laboratory, MPI-FKF and CNRS, B.P. 166, 38042 Grenoble Cedex 9, France}
\affiliation{L. D. Landau Institute for Theoretical Physics, Academy of Sciences of Russia,
 117940 Moscow V-334, Russia}

\date{\today}

\begin{abstract}
The Fr\"{o}hlich interaction is one of the main electron-phonon intrinsic interactions in polar materials
originating from the coupling of one itinerant electron with the macroscopic electric field generated by
any longitudinal optical (LO) phonon. Infra-red magneto-absorption measurements of doped GaAs quantum wells
structures have been carried out in order to test the concept of Fr\"{o}hlich interaction and polaron mass
in such systems. These new experimental results lead to question the validity of this concept in a real system.

\end{abstract}


\maketitle

Based on the Fr\"{o}hlich interaction, polaronic effects were early  studied  by Lee and Pines
\cite {Lee} and also by Feynman \cite{Feynman} on theoretical grounds:  simple consequences were derived
leading to the  concept of the polaron mass, and reproduced in many text books \cite{Kittel}.
This conceptual interaction also attracted much attention, on the experimental side, in either three or
quasi two-dimensional (3D or Q2D) semiconductors \cite{Devreese}  which were model systems to study the effect.
For the simplest case of a Q2D semiconductor, the single-particle energy spectrum of electrons of mass $m^{*}$,
in  a perpendicular magnetic field $B_{\bot}$, consists of equally spaced ($\hbar\omega_{c\bot} = eB_{\bot} /m^{*}$)
energy states known as Landau levels (LL) labelled with an integer index n going from 0 to infinity:
one direct consequence of the polaronic concept is the prediction of the resonant magneto-polaron coupling
(RMPC) which anticipates a non linear field variation of the  cyclotron resonance (CR) energy, $ \hbar\omega_{c}$,
 when $\hbar\omega_{c}\simeq\hbar\omega_{LO}$, with a pinning of the CR \textit{above} $\hbar\omega_{LO}$.
 Such  experimental evidence is quite difficult to obtain because
the range of energies concerned coincides with that of the strong absorption of the phonon band (Restrahlen)
of the material, therefore obscuring the data. The theory itself raises also fundamental questions because,
 in real materials, one always has many electrons which are known to oscillate in phase as plasmons
developing in turn a macroscopic electric field which couples to that of the LO phonon, generating hybrid
plasmon-phonon modes. What really happens? Does the polaronic concept
remains valid?  To our knowledge there is no quantum mechanical treatment of the hybrid modes and therefore it
is difficult to answer these questions on theoretical grounds.

Experiments performed in suitable conditions are however able to bring new light in that matter.
Overcoming the problem of the Restrahlen band  is indeed technically very difficult
in 3D semiconductors. In that respect the Q2D structures, grown by molecular beam epitaxy
(MBE), are more appealing because, besides the fact that, with the remote doping technique, carriers can
 remain quasi-free at the lowest temperatures, one can remove, by
selective chemical etching, the whole MBE structure from the substrate (lift-off process). It is then possible to
deposit it on an infra-red transparent substrate, like Si for instance, which is wedged with an angle of few
degrees to avoid interference effects. This structure is composed of a single doped GaAs quantum well (QW) of
width $L$ and areal density $n_{S}$ sandwiched between two GaAs-AlAs superlattices with the Si-n type doping
performed symmetrically on both sides of the QW \cite{Friedland}. This enables to grow structures of quasi 2D
electron gas (2DEG) combining high density and high mobility. We already reported  results \cite{Poulter}
on such a sample with $L = 10$ $ nm, n_{S}= 12\times 10^{11}cm^{-2}$ \cite{comment}. The same process has been applied
to new samples with similar structures but with $ L=13$ $nm$ and various lower doping levels. Magneto infrared
transmission measurements are done in an absolute way for any
fixed magnetic field $B$: this is achieved with a rotating sample holder allowing to measure  reference
spectra in the same conditions on a parent wedged piece of Si substrate. With the $\overrightarrow{k}$  vector of the
incoming light parallel to $\overrightarrow{B}$, the growth axis of the sample was maintained either parallel to this
direction (perpendicular Faraday (PF) configuration) or tilted by an angle $\theta$ with respect to it
(tilted  Faraday (TF) configuration). The light was guided through an over-sized guide pipe ending by a cone to minimize
the divergence of the beam. The interpretation of the experimental results is done by comparing them with a simulation
of the multi-layer transmission by an appropriate model.
This is essential because, in the frequency range of interest, the spectra can be distorted by dielectric
interference effects independent of any specific electron-phonon interaction.

The previous study   \cite{Poulter} reported on  PF configuration  measurements where no sign
of RMPC was observed. This was not
necessary a proof that the effect was not existing because it could have been screened by the high value of $n_{S}$ :
indeed theoretical simulations \cite{Wu}, including the RMPC, of data obtained with such high density QW was
still showing a small interaction.   New experiments with a lower $n_{S}$, for which this simulation provided clear effects,
 were necessary to check the concept of the RMPC.
 The doping level of the new lift-off samples ranges between
  $ n_{S} = 7.0 \times 10^{11}cm^{-2}$ (sample 1200) and  $n_{S} =  9.0\times 10^{11}cm^{-2} $(sample 1211) with
   respective mobilities of $\mu_{DC}= 260\, m^{2}/V/sec$ and $\mu_{DC}= 280\, m^{2}/V/sec$
  \cite{comment}.
 All experimental  results
reported here have been obtained at a fixed temperature of 1.8 K. We will concentrate  on the results obtained for
$B_{\bot}$  around 20 T for which the CR span the Restrahlen band of GaAs. For both samples, in that range of fields,
the filling factor of the structure $\nu =n_{S}h/(eB)$ remains essentially lower than 2 and, in that case,
non-parabolicity (NP) effects can be ignored.

Typical spectra, for  sample 1200, are displayed in  Fig.1.
The absolute transmission (AT) at $B = 0 T$ and $\theta = 60^{o}$ is shown in Fig. 1a. The strong absorption
features are those related to the TO phonons of GaAs and AlAs whereas the weak ones are those related to
the LO modes which are known to become active in absorption for thin slabs \cite{Berreman} in the TF configuration.
They clearly originate from the layers of the superlattices surrounding the doped QW. They are slab modes and give
a lower bound of the energy of LO modes in the whole structure. In Fig. 1b, the absolute transmission of the sample
in a TF configuration is displayed for $\theta = 20^{o}$ and $B_{\bot} = 20.7 T$. In practice with the available
magnetic fields, we are led to work with angles where the LO phonons are less visible but they are still clearly
present as indicated by the vertical dashed lines.
\begin{figure}
\includegraphics*[width=0.7\columnwidth]{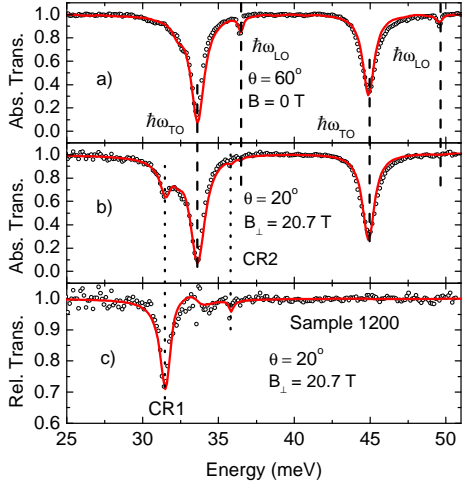}
\caption{\label{Fig.1} Experimental transmission spectra (empty dots) of sample 1200.  a) absolute transmission
for $B = 0\,T$ and $\theta = 60^{o}$; b) and c) absolute transmission and relative transmission respectively
for $B_{\bot} = 20.7\,T$ and $ \theta = 20^{o}$. The continuous red curves are the fit of data with the model
described in the text.}
 \end{figure}
Two  CR transitions, labelled CR1  and CR2 (vertical dotted lines),
are now present, the CR2 feature being, for that field, at a \textit{lower} energy than $\hbar\omega_{LO}$(GaAs).
In Fig. 1c we show the relative transmission (RT) obtained by dividing the AT spectrum of Fig. 1b by the AT spectrum
 at $B = 0 T$ for the same angle. The RT spectra are a combination of four experimental ones and therefore the noise
 is more important. However the only remaining features  are those related to the contribution of the 2DEG.
 The energies of the CR1 and CR2 structures for each sample, traced as a function of $B_{\bot}$, are displayed in Fig. 2
  for different angles.
\begin{figure}
\includegraphics*[width=0.7\columnwidth]{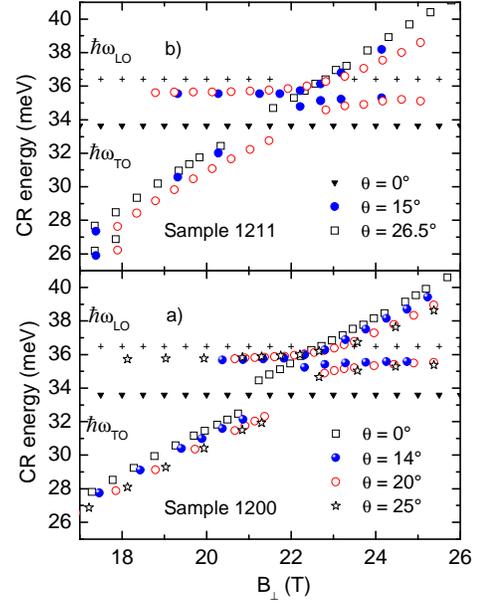}
\caption{\label{Fig.2} Observed cyclotron energies for sample 1200 (a) and sample 1211 (b) for different tilt angles  .
The crosses and black down triangles are the measured LO and TO energies respectively and the  empty squares the
cyclotron energies in the PF configuration. Other points are the observed CR energies in the TF
configuration: (a) full dots : $\theta = 14^{o}$, empty dots : $\theta= 20^{o}$, stars : $\theta = 25^{o}$ and
(b) full dots : $\theta= 15^{o}$, empty dots : $\theta= 26.5^{o}$.
Results in (b) for $B_{\bot}$  lower than $18 \,T$ correspond to filling factors $>$ 2 when non parabolicity effects are
apparent.}
 \end{figure}
Whereas in the PF configuration (empty black squares), no sign of interaction is observed, in the TF
configuration a pronounced anticrossing of the CR1 and CR2 components occurs, which increases with $\theta $,
one of the components  remaining pinned at an energy of $35.6 \pm 0.05 $ meV for the sample 1200 and
$35.5 \pm 0.05 $ meV for the sample 1211, both energies being \textit{lower} than $\hbar\omega_{LO} $(GaAs).
Other samples with
different densities of electrons ($n_{S} = 5.8 \times 10^{11} cm^{-2} $ and $n_{S} = 8.8 \times 10^{11} cm^{-2} $ and
the same $L $ of 13 nm or with a smaller $L $ of 10 nm and $n_{S} = 12 \times 10^{11} cm^{-2} $, have also been
investigated and  demonstrate exactly the same qualitative features.

To interpret these results, one has to calculate  the transmission of the whole structure. This  requires a full
derivation of the multilayer structure transmission in the TF configuration, a rather complicated problem which we
have succeeded to solve in some specific cases and which will be detailed in a further publication \cite{Bychkov}.
We  just  here  outline the principles. We used a standard approach which consists in evaluating for each layer N,
with  a dielectric tensor $\ten{\varepsilon}_{N}(\omega)$, the transfer matrix $M_{N}$
after finding the appropriate modes of propagation of light $\overrightarrow{k}^{i}_{N}$  inside the layer, by solving the
Maxwell equations. Taking the z-axis perpendicular to the layer, and defining the y-z plane as the plane of
incidence which also contains $\overrightarrow{B}$, the angle $\theta$  is defined as the angle
($\overrightarrow{z},\overrightarrow{B}$). There are four modes $k^{i}_{z,N}$ for each layer.
The $4\times4$ transfer matrix is built by writing the conservation of the
tangential components of the electric and magnetic fields of the light at each interface of the layer. If
$\ten{\varepsilon}_{N}(\omega)$ is diagonal, like in  layers which do not contain free electrons
(or at  $B = 0 T$) the transfer matrix  decomposes into two blocks of $2\times2$ matrices related to
the TE and TM modes respectively. This result is well known and has been used \cite{Sciacca}  to measure the LO phonons
frequencies of semiconductors in a way similar to the results reproduced in Fig. 1a. For the doped QW,
in the TF configuration, all elements of $\ten{\varepsilon}_{QW}(\omega)$ and therefore of the transfer matrix
 are non zero and the TE and TM modes are mixed.  The resulting total transfer matrix  has the
same properties and a special treatment \cite{Bychkov} needs to be developed to extract the transmission of the
structure. This can be done as soon as
the dielectric tensor of the QW in the TF configuration has been defined.

In the PF configuration, for any finite value of $B$, the z-part of the 2DEG  wave function is decoupled from
the x-y part. When $\theta\neq0$,
this decoupling is no longer valid  and as we shall see below is indeed responsible for the observed
anticrossing. One has then to reproduce the mixing of the z and x-y part of the 2DEG wavefunction with the best accuracy.
For the square QW which is under study, this mixing can only be evaluated in perturbation theory \cite{Bastard}. On the
other hand,  if we assume the z-confining potential to be parabolic, with a characteristic energy $\hbar\Omega$
separating the electric subbands, the coupling can be evaluated analytically at all orders. Of course the strength of
the coupling is modified, the dipole oscillator strengths along the z-axis being different between a square QW and a
parabolic one. It is this model, based on the Drude formalism, that we have used to evaluate, neglecting retardation
effects,  all components  of $\ten{\varepsilon}_{QW}(\omega)$ and then simulate the transmission. Before discussing
these simulations it is instructive to  look at the structure of two characteristic components
of $\ten{\varepsilon}_{QW}(\omega)$.
The real part of $\varepsilon_{xx}$ and $\varepsilon_{zz}$  are expressed, ignoring damping effects,
as \cite{Bychkov} :
\begin{eqnarray}
\varepsilon_{xx}= \varepsilon_{L}-\frac{\omega_{p}^{2}(\omega^{2}-\Omega^{2})}{(\omega^{2}-\omega_{1}^{2})
(\omega^{2}-\omega_{2}^{2})}\\
\varepsilon_{zz}= \varepsilon_{L}-\frac{\omega_{p}^{2}(\omega^{2}-\omega^{2}_{c\bot})}
{(\omega^{2}-\omega_{1}^{2})(\omega^{2}-\omega_{2}^{2})}
\end{eqnarray}
where $\varepsilon_{L}$ is the contribution of the lattice, $\omega_{p}^{2}= 4\pi e^{2}n_{S}/(Lm^{*})$
the square of the plasma frequency, $\omega_{1}^{2}$  and $\omega_{2}^{2}$  are the solutions of the equation:
\begin{equation}
\omega^{4}- (\Omega^{2}+ \omega_{c}^{2})\omega^{2}+ \omega_{c\bot}^{2}\Omega^{2}=0
\end{equation}
with $\omega_{c} = eB/(\hbar m^{*})$. When $\Omega >> \omega_{c\bot}$, $\omega_{1}\rightarrow\omega_{c\bot}$ and
$\omega_{2}\rightarrow\Omega$. In the present case, however, the exact values for the poles are introduced in the
calculation. This explains why in Fig. 2, far from the anticrossing region the energies of the CR lines do not
coincide with the values obtained in the PF configuration.

Different fitting steps have to be performed. We first have to fit $\varepsilon_{L}$ with standard values as
already done \cite{Poulter} and shown in Fig. 1a (red continuous lines). In fact, in the region of $\hbar\omega_{TO}$(GaAs),
interference effects, between phonon and CR absorptions become important and  this is the reason
why we prefer to interpret the relative transmission spectra (Fig. 1c and Fig. 3) for which these spurious effects
are minimized. They are however not eliminated and, for this reason,  we ignore in the fitting process a region of
$\pm 0.5$ meV around $\hbar\omega_{TO}$(GaAs), sketched as the  hatched region in Fig. 3. In this figure,  experimental data
(empty dots) and their simulations (red continuous curves) for sample 1200  are displayed for $\theta =25^{o}$.
\begin{figure}
\includegraphics*[width=0.7\columnwidth]{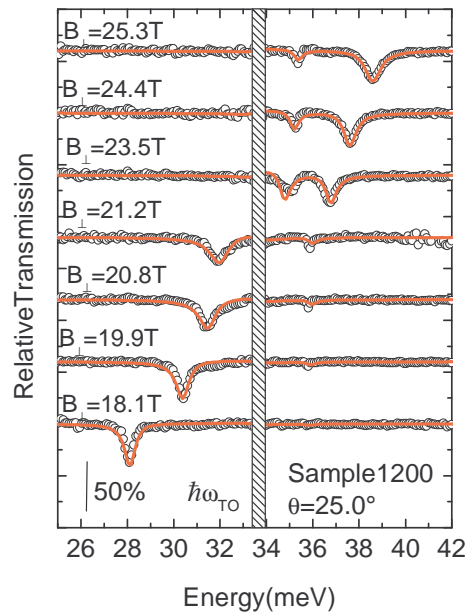}
\caption{\label{Fig.3} Fit of the relative transmission curves for sample 1200 at $\theta = 25^{o}$. The curves are shifted for clarity.
Empty dots are experimental points and full continuous curves the corresponding calculated relative transmission spectra.
 The hatched region corresponds to that of $\hbar\omega_{TO}$(GaAs) (see text).}
 \end{figure}
There are different fitting parameters to adjust: $n_{S}$, $m^{*}$, $\tau_{CR}$ , $\Omega$ and $\theta$, $\tau_{CR}$
being the scattering time entering the Drude model . In the PF configuration, $n_{S}$ is fitted  at lower fields
with the model extended for NP effects \cite{Bychkov} together with $m^{*}$  and $\tau_{CR}$. Outside the
energy region around $\hbar\omega_{TO}$(GaAs) where
there are signs of some interaction (at least for the sample 1200) which are not discussed here, the values
of $m^{*}$ fitted correspond quite well to the expected variation induced by NP effects \cite{Hermann}
 whereas the mobility $\mu_{CR} = e \tau_{CR}/m^{*}$ ranges around $150\, m^{2}/V/sec$ revealing the good quality of
 the samples.  The angle $\theta$ has also to be fitted because, for technical reasons, the Si substrate being
 wedged, $\theta$ can differ by a few degrees from the mechanical angle. It is fitted by setting the $m^{*}$
 value constant for a given value of $B_{\bot}$. This is not completely independent of the value of $\Omega$
 but the loop  easily converges.
 The value of  $B_{\bot}$ indicated in the figures are calculated values of $B\times cos(\theta)$. Finally $\Omega$
 is fitted  in such a way the splitting of the CR1-CR2 lines can be best reproduced. We impose also, in
 the fitting process, the value of $\Omega$ to be constant for all angles and QW having the same width $L$.
 As an example, the results, shown in Fig. 3 for sample 1200,  are obtained for a value of $\Omega = 63$ meV
 (with an uncertainty of $\pm 3$ meV).
 These  results clearly demonstrate the anticrossing features observed experimentally. The fit  reveals some slight
 deficiency which is sample dependent but the overall agreement is surprisingly good.  It is interesting to note
  that if one calculates, at $B = 0 T$, the zeros of $\varepsilon_{zz}$ (Eq. 2) with the fitted value
 of $\Omega$, one finds for the low energy modes 35.62 meV for the sample 1200 and 35.49 meV for the sample 1211,
 values which agree very well with the experimental findings. These modes are the so called plasmon-phonon-intersubband
 (PPI) modes \cite{Pinczuk}. Therefore the observed anticrossing occurs between the CR mode and this hybrid PPI mode.
 They are coupled by symmetry but not by any  \textit{specific electron-phonon interaction}. We think, though approximate,
  the model keeps all the physical and symmetry aspects of the problem. One direct consequence is that,
   when $\Omega$ increases, the
  oscillator strength of the PPI-like mode rapidly decreases to zero \cite{Bychkov}. Therefore, for a pure 2DEG, there
  is no   longer any interaction whatever is the angle. The data in the PF and TF configurations do not show any
  sign of interaction related to the  Fr\"{o}hlich coupling: we are therefore led to conclude that
  the concept of polaronic mass, related to  this interaction, is no longer effective.

All the discussion is based on the assumption that we are dealing with carriers for which a plasma frequency can be
defined. It is known that for bound electrons, like the ones found in neutral shallow impurities, the $1s-2p^{+}$ absorption
reveals \cite{Huant} the Fr\"{o}hlich interaction around the energy of the LO phonon in GaAs but then the mass of
such an electron is not defined. The same situation holds for electrons in  quantum dots \cite{Hameau}. It is also true
that there is still some electron-phonon interaction present in such  system : indeed, around the TO frequency, there
is a clear sign of interaction especially in sample 1200. This is an interaction which  may be driven by
the mechanism of deformation potential which is at the origin, for instance, of  the Raman response of such structures.

In conclusion, we have performed careful magneto-optical measurements of the infra-red absorption related to quasi
two-dimensional electron gas, with different densities, in a range of magnetic fields which allows the cyclotron
resonance to span the optical  range of phonon energies. All singularities observed can be explained quantitatively
without any electron-phonon coupling of the Fr\"{o}hlich type. We argue that this support the idea that the concept
of the Fr\"{o}hlich polaron mass has to be re-examined in a real material.

The GHMFL is "Laboratoire conventionn\'{e} à l'UJF et l'INPG de Grenoble". The work presented here has been supported
in part by the European Commission through the Grant HPRI-CT-1999-00030. Yu. B. acknowledges the support of
the grant No. RFFI-03-02-16012.

\bibliography{basename of .bib file}

\begin{references}
\bibitem{comment} The carrier densities reported in this paper, for all samples, is obtained from the fit of data
with the model detailed in the following including non-parabolicity effects. They are systematically lower by 6
to 8 per cent with respect to the value obtained from transport data on the parent non lift-off sample.
The values of mobilities given here are however those obtained from these data.
\end{references}

\end{document}